\documentstyle[preprint,aps,psfig,prb]{revtex}
\tighten
\begin{document}
\draft
\date{\today}
\title{Implementation of the Projector Augmented Wave LDA+U Method:\
Application to the Electronic Structure of NiO}
\author{O. Bengone$^{1,2}$, M. Alouani$^1$, P. Bl\"ochl$^3$, and J. Hugel$^2$}
\address{$^1$Institut de Physique et Chimie des Mat\'eriaux de
Strasbourg (IPCMS), 23 rue du Loess, F-67037 Strasbourg Cedex, France}
\address{$^2$Univesit\'e de Metz, IPC-LPLI Groupe de Structure Electronique
des Milieux denses, 1 bd Arago, F-57078 Metz Cedex 3, France}
\address{$^3$IBM Research, Zurich Research Laboratory, CH-8803
R\"uschlikon, Switzerland}
\maketitle
\begin{abstract}
  The so-called local density approximation plus the multi-orbital
  mean-field Hubbard model (LDA+U) has been implemented within the
  all-electron projector augmented-wave method (PAW), and
  then used to compute the insulating antiferromagnetic ground
  state of NiO and its optical properties. The electronic and optical
  properties have been investigated as a function of the Coulomb
  repulsion parameter $U$. We find that the value obtained from
  constrained LDA ($U=8$ eV) is not the best possible choice, whereas
   an intermediate value ($U=5$ eV) reproduces the experimental
  magnetic moment and optical properties satisfactorily.  At
  intermediate $U$, the nature of the band gap is a mixture of charge
  transfer and Mott-Hubbard type, and becomes almost purely of the charge-transfer
  type at higher values of $U$. This  is due to the
  enhancement of the oxygen $2p$ states near the top of the valence
  states with increasing $U$ value.
\end{abstract}
\section{Introduction}
\label{sec:one}
For many materials, the density functional theory (DFT)\cite{kohn1} in the
local-spin-density approximation (LSDA)\cite{kohn2} provides a good description
of their ground-state properties.  However, problems arise when the DFT-LSDA
approach is applied to materials with ions that contain incomplete $d$- or
$f$-shells, such as transition-metal oxides or heavy fermion systems. For
example, most transition-metal oxides are wide-gap antiferromagnetic
insulators,\cite{powell,messick,sawa1,hufner1,hufner2,hufner3,elp2} and the
DFT-LSDA predicts them to be either metals (FeO and CoO) or small-gap
semiconductors (MnO and NiO).\cite{tera}  The failure of the DFT-LSDA can be
traced to the mean-field character of the Kohn-Sham equations as well as to the
poor description of strong correlation effects within the homogeneous electron
gas. The strong correlation effects are responsible for the breakdown of the
DFT-LSDA description of the electronic structure of these compounds.  In order
to provide a better description of these effects, the Mott-Hubbard
picture has been introduced.\cite{mott,hubbard}

In the Mott-Hubbard picture of NiO, the $d$-$d$ Coulomb interaction splits the
Ni $d$ sub-bands into the so-called lower and upper Hubbard bands.  The upper
Hubbard band has mostly Ni $3d^9$ character, while the top of the valence band
is of $3d^8$ character, leading to a Mott-Hubbard gap of $d$-$d$ type. However,
O $1s$ x-ray absorption\cite{kuiper} as well as x-ray photoemission and
bremsstrahlung isochromat spectroscopies\cite{elp2} on Li$_x$Ni$_{1-x}$O have
shown that the additional hole has mainly oxygen character.  In contrast to the
Mott-Hubbard model, the energy band gap caused by the Ni $3d$ correlations is
therefore of the charge-transfer type between the occupied oxygen $2p$ and the
Ni $3d$ empty states.

On the other hand, localized approaches,
\cite{fuji1,janssen,okada,zheng,benesh,tanaka2,zaanen} such as the local
cluster scheme based on the configuration interaction method or the Anderson
impurity model, in which transition-metal ions are treated like an impurity in
an oxygen $2p$ host, predict a well-defined band gap of 5.0 eV. However, the
oxygen $2p$ band dispersion observed in angle-resolved photoemission
spectroscopy\cite{shen1} cannot be described by these methods because the
lattice effects are neglected.

Several attempts have been made to include the missing correlation effects in
DFT-LSDA. The generalized gradient approximation (GGA),\cite{perdew92} which
takes into account the radial and angular gradient corrections, can only open a
small band gap.\cite{dutek}  The self-interaction correction
(SIC)\cite{svane90,temmerman93,fujiwara} eliminates the spurious interaction of
an electron with itself from the conventional DFT-LSDA method. Compared to
LSDA, the band gap and the magnetic moments are significantly increased.
However, the band gap still is too small, and the SIC-LSDA method predicts a
larger energy band gap for NiO than for FeO and CoO, in contradiction to
experiment.\cite{powell}  The crystal-field orbital polarization introduced by
Norman \cite{norman91} to determine the magnetism and insulating band gap of
transition-metal oxides is promising but underestimates both the spin magnetic
moment and the band gap.

Another promising approach for correlated materials is the so-called local
density approximation (LDA) plus the multi-orbital mean-field Hubbard model
(LDA+U)\cite{anisi1,anisi3,pickett,Surfnio,shick,solov,liechten} which includes
the on-site Coulomb interaction in the LSDA Hamiltonian. After adding the
on-site Coulomb interaction to the LSDA Hamiltonian, the potential becomes
spin- {\it and} orbital-dependent. Because a larger energy cost is associated
with fluctuations of the $d$-occupancy, the orbital-dependent potential reduces
the fluctuations of the $d$-occupancy, resulting in a better justification of a
mean-field approach.  LDA+U, although it is a mean-field approach, has the
advantage of describing both the chemical bonding and the electron-electron
interaction.

The question regarding the best value for the Coulomb repulsion parameter $U$
is, however, still under debate.  The $U$ parameter for NiO obtained from a
constrained LDA calculation is about 7 to 8 eV,  and this is the value
generally used in LDA+U calculations.  A similar value of $U$ has been obtained
from a constrained LDA calculation for bulk Fe,\cite{anisi3} even though a much
smaller value had been expected because of the metallic screening in Fe. The
authors argued\cite{anisi3} that the higher value could be an artifact due to
the poor screening within the atomic sphere approximation (ASA), and that
within a full-potential calculation a much smaller value of less than 4 eV
would be expected.  In contrast, an unpublished full-potential linear
muffin-tin orbital method calculation by Alouani and Wills\cite{awunpub}
clearly shows that the value of $U$ for bulk Fe is even slightly larger than
the ASA value. Furthermore, Bulut, Scalapino, and White\cite{bulut} showed that
the renormalization of the Coulomb interaction depends on the type of model
used. As LDA is not a diagrammatic method, it is not known which type of
renormalization is the most appropriate for the LDA+U model.

In this work we shed light on this problem by treating the Coulomb repulsion
parameter $U$ as adjustable parameter, and by investigating how the electronic
and optical properties depend on its value.  We show that for an intermediate
value of $U = 5$ eV, good agreement with the measured ground-state
antiferromagnetic magnetic moment and optical properties is obtained.  We also
show that the O $2p$ character near the top of the valence states is enhanced
for a larger value of $U$.  Our calculation seems to indicate that the nature
of the band gap at intermediate $U$ is a mixture of charge transfer and
Mott-Hubbard type, and that it becomes almost purely of the charge-transfer
type for higher values of $U$.

Our calculations are based on the projector augmented wave (PAW)
method,\cite{paw} an efficient all-electron method without shape approximations
on the potential or electron density to avoid uncertainties due to the ASA
approach. Based on a Car-Parrinello-like formalism,\cite{car85} the PAW method
allows complex relaxations and dynamical properties in strongly correlated
systems to be studied. Our implementation of LDA+U within the PAW method is
described in detail.  Furthermore we discuss possible extensions of the
existing method that will enhance its applicability.

The paper is organized as follows:\ In Section \ref{sec:two} we present those
aspects of the PAW formalism that are needed for the implementation of the
LDA+U method.  In Section \ref{sec:three} we present and discuss the LSDA and
LDA+U ground-state properties of NiO, and in Section \ref{sec:four} we study the
optical properties of NiO, namely the imaginary part of the dielectric
function, and compare the results to experiment.

\section{Formalism}
\label{sec:two}
\subsection{PAW method}

The PAW method developed by one of us \cite{paw} combines ideas of the
pseudo-potential (PP) and the linear augmented plane wave (LAPW) methods. It is
applicable to all elements of the periodic table. The nodal behavior of the
wave function is correctly described and, as in the PP method, the forces on
the ions are easily expressed.

In the PAW method, the all-electron (AE) crystal wave function is constructed
from a pseudo (PS) wave function and atom-like functions localized near the
nuclei.  The PS wave function $|\tilde\Psi\rangle$ coincides with the crystal
AE wave function $|\Psi\rangle$ in the interstitial region, i.e.\ outside the
atomic regions.  Inside the atomic regions $\Omega_t$, called augmentation
regions, the wave function is almost atom-like because the effect of the
surrounding crystal is small.  Therefore, a natural choice is to use solutions
$|\phi_\Lambda\rangle$ of Schr\"odinger's equation for the isolated atom,
the so-called AE partial waves, as a basis set for the augmentation region. Here
$\Lambda=\{t,\alpha,\ell,m\}$ is a global index for the atom $t$, the angular
momentum $\ell$, the magnetic quantum number $m$, and the index
$\alpha$,
the energy for which Schr\"odinger's equation is solved.

To link the expansion in atom-like functions near the nuclei to the PS wave
function, we introduce a set of auxiliary functions
$|\tilde\phi_\Lambda\rangle$, so-called PS partial waves, which are centered on
the atom and coincide per construction with the corresponding AE partial waves
$|\phi_\Lambda\rangle$ outside their augmentation regions:
\begin{equation}
\phi_{\Lambda}({\bf r})=\tilde\phi_\Lambda({\bf r}) \qquad \textrm{for} \  r
\notin \Omega_t \, .
\end{equation}

The coefficients $c_\Lambda$ of the expansions in AE and PS partial waves are
chosen such that the PS partial wave expansion $\sum_\Lambda
|\tilde\phi_\Lambda\rangle c_\Lambda$ cancels out the PS wave function
$|\tilde\Psi\rangle$ inside the augmentation region.  For this purpose we
introduce so-called projector functions $\langle\tilde{p}|$ such that
\begin{equation}
\sum_\Lambda |\tilde\phi_\Lambda\rangle\langle\tilde{p}_\Lambda|=1 \, ,
\label{eq:pawident}
\end{equation}
and therefore
\begin{equation}
|\tilde\Psi\rangle=\sum_\Lambda |\tilde\phi_\Lambda\rangle
\langle\tilde{p}_\Lambda|\tilde\Psi\rangle
\label{eq:psexpand}
\end{equation}
for the Hilbert space spanned by the PS partial waves
$|\tilde\phi_\Lambda\rangle$. Thus we identify the expansion coefficients with
$c_\Lambda=\langle\tilde{p}_\Lambda|\tilde\Psi\rangle$. Equation
(\ref{eq:pawident}) results in the biorthogonality condition
\begin{equation}
\langle\tilde{p}_\Lambda|\tilde\phi_{\Lambda'}\rangle=\delta_{\Lambda,\Lambda'}
\end{equation}
for the projector functions, which moreover are chosen to be localized within
the corresponding augmentation region.

With these conditions, the AE Bloch wave function $\Psi({\bf r})$ can
be obtained from the PS wave function $\tilde\Psi({\bf r}) $ as
\begin{equation}
\Psi({\bf r})=\tilde\Psi({\bf r}) + \sum_{\Lambda} [\phi_{\Lambda}({\bf r})
-\tilde\phi_\Lambda({\bf r})] \langle\tilde{p}_\Lambda|\tilde\Psi\rangle \, .
\label{eq_psi_paw}
\end{equation}
The first term represents the PS wave function defined over the entire space,
which is equal to the AE wave function in the interstitial region, and which is
expanded in plane waves.  The second term is the AE partial wave expansion,
which describes the correct nodal behavior of the wave function in the
augmentation region $\Omega_R^t$ ($r \leq r_c^t$).  The third term eliminates
the spurious contribution of the PS wave function in the augmentation region.

Note that Eq.\ (\ref{eq:psexpand}) holds only approximately if the set of PS
partial waves is not entirely within the augmentation regions. However, this
has the advantage that only those contributions of the PS wave function will be
removed that are also replaced by AE partial waves. As a result, the AE wave
function converges rapidly with the number of partial waves used and, moreover,
it is continuous and differentiable for every truncation of the partial-wave
expansion.

Expectation values of any sufficiently local operator $A$ are obtained as
\begin{equation}
\langle\Psi|A|\Psi\rangle=
\langle\tilde\Psi|A|\tilde\Psi\rangle+\sum_{\Lambda,\Lambda'}
\langle\tilde\Psi|\tilde{p}_\Lambda\rangle
\Bigl(
\langle\phi_\Lambda|A|\phi_{\Lambda'}\rangle
-\langle\tilde\phi_\Lambda|A|\tilde\phi_{\Lambda'}\rangle
\Bigr)
\langle\tilde{p}_{\Lambda'}|\tilde\Psi\rangle \, .
\label{eq:pawoperator}
\end{equation}
Note that the double sum is diagonal in the site indices $t,t'$.
This equation is exact for a complete set of PS partial waves and rapidly
attains the converged result if incomplete.  The PAW method provides the
freedom to represent a zero operator in the form
\begin{equation}
0=
\langle\tilde\Psi|B|\tilde\Psi\rangle-\sum_{\Lambda,\Lambda'}
\langle\tilde\Psi|\tilde{p}_\Lambda\rangle
\langle\tilde\phi_\Lambda|B|\tilde\phi_{\Lambda'}\rangle
\langle\tilde{p}_{\Lambda'}|\tilde\Psi\rangle
\label{eq:pawzero}
\end{equation}
by any operator $B$ entirely localized within the augmentation region. Equation
(\ref{eq:pawzero}) has the same range of validity as Eq.\
(\ref{eq:pawoperator}) does and allows a further acceleration of the
convergence by using a well-chosen operator $B$ and adding the corresponding
zero operator to the expression for the expectation value.

\subsection{The LDA+U total energy functional}
For transition-metal oxides, the $d$-orbitals are well localized and keep a
strong atom-like character. Even though LDA provides a good approximation for
the average Coulomb energy of the $d$-$d$ interactions, it fails to describe
correctly the strong Coulomb and exchange interaction between electrons in the
same $d$-shell.  The main intention of LDA+U is to identify these atomic
orbitals and to describe their electronic interactions as strongly correlated
states. The other orbitals are delocalized and considered to be properly
described by the LDA.  The procedure is to eliminate the averaged LDA energy
contribution of these atom-like orbitals from the LDA total energy functional
$E^{\rm LDA}$, and to add an orbital- and spin-dependent correction. The total
energy within the LDA+U method then has the form
\begin{eqnarray}
E&=&E^{\rm LDA} + \frac{1}{2} \sum_{t,\sigma}
\sum_{i,j,k,l}\langle\chi^t_i;\chi^t_k| V_{ee}|\chi^t_j;\chi^t_l\rangle
n^{t,\sigma}_{i,j}n^{t,-\sigma}_{k,l} \nonumber\\
&+& \frac{1}{2} \sum_{t,\sigma}\sum_{i,j,k,l}\Bigl(
\langle\chi^t_i;\chi^t_k|V_{ee}|\chi^t_j;\chi^t_l\rangle
-\langle\chi^t_i;\chi^t_k|V_{ee}|\chi^t_l;\chi^t_j\rangle \Bigr)
n^{t,\sigma}_{i,j}n^{t,\sigma}_{k,l} \nonumber\\ &-&\sum_{t}\left [
\frac{1}{2}U
\sum_{\sigma,\sigma'}N^{t,\sigma}(N^{t,\sigma'}-\delta_{\sigma,\sigma'})
-\frac{1}{2}J\sum_\sigma N^{t,\sigma}(N^{t,\sigma}-1)\right ] \, ,
\label{Eq:Etot}
\end{eqnarray}
where $N^{t,\sigma}=\sum_{i}n^{t,\sigma}_{i,i}$ is the average
occupation of the $d$-shell for each spin direction as obtained from
the $d$-orbital occupancies $n^{t,\sigma}_{i,j}$. $U$ and $J$ are the
Coulomb self-energy and the exchange parameter, respectively.  The
expressions $\langle\chi^t_i;\chi^t_k|V_{ee}|\chi^t_j;\chi^t_l\rangle$
are the four-center matrix elements of the screened Coulomb interaction
$V_{ee}$. An additional requirement of the LDA+U approach is that the
additional energy is applied only to the valence electrons, which are
re-optimized while constrained to remain orthogonal to the core states.

\subsection{Orbital occupations}

The rationale behind the definition of the orbital occupations is to
project the density matrix onto one consisting of a restricted set
of localized orbitals $|\chi^t_{i}\rangle$, where each orbital index
$i$ stands for an angular momentum and $t$ for an atomic site.
Here we focus in particular on the $d$-orbitals of Ni.

The projected density matrix operator has the form
\begin{equation}
\bar{\rho}^\sigma=\sum_{i,j}|\chi^t_i\rangle
n^{t,\sigma}_{i,j}\langle\chi^t_j| \, .
\end{equation}
The density matrix elements $n^{t}_{i,j}$ in the localized basis set
can be obtained from the condition that the expectation values of a set
of projection operators $P_s$ are optimally reproduced by the localized
basis set, i.e.\ the weighted square deviation
\begin{eqnarray*}
F&=&\sum_s w_s \left[\sum_{n, {\bf k}} \langle\Psi^{{\bf
k}}_n|P^t_s|\Psi^{{\bf k},\sigma}_n\rangle -\sum_{i,j}
n^{t,\sigma}_{i,j}\langle\chi^t_i|P^t_s|\chi^t_j\rangle\right]^2
\end{eqnarray*}
of the projections should be minimized.

Here we choose the projection operators acting on the Ni $d$-orbitals
in analogy to previous implementations\cite{shick} as
\begin{equation}
P^t_{m,m'}({\bf r},{\bf r'})=\theta_{\Omega_t}({\bf r}) \delta(|{\bf
r'-R_t}|-|{\bf r-R_t}|)Y_{d,m}^t({\bf r}) Y_{d,m'}^*({\bf r'}) \, ,
\label{eq:projector}
\end{equation}
where the site index $t$ refers to a particular Ni site, and
$|Y_{d,m}^t\rangle $ is the spherical harmonic centered at site $t$.
The step functions $\theta_{\Omega_t}(r)$ are unity for $|r-R_t|<r^t_c$
and zero otherwise. $Y_{d,m}$ are the spherical harmonics for $\ell=2$.
(Note that $\langle\chi|P|\chi'\rangle=\int d{\bf r}\int d{\bf r'}
\chi({\bf r})P({\bf r},{\bf r'})\chi'({\bf
  r'})$.)  The radius $r^t_c$ for Ni corresponds to an atomic sphere
radius, and we have chosen $r_c^t=2.1$ a$_0$.
%
The localized orbitals $|\chi^t_m\rangle$ have been chosen to be
identical to those of the spheridized, non-spin-polarized atoms. We
used an atom in the $3d^84s^2$ configuration of Ni. For such a
localized Ni $d$-orbital $|\chi^t_m\rangle$ with magnetic quantum
number $m$, we obtain
\begin{equation}
I=\langle\chi^t_m|P^t_{m,m}|\chi^t_m\rangle=
\langle\chi^t_m|\theta_{\Omega_t}(r)|\chi^t_m\rangle \, ,
\end{equation}
independent of $m$ and $t$.
Thus we obtain the orbital occupations
\begin{equation}
n^{t,\sigma}_{m,m'}=\frac{1}{I}\sum_{n,{\bf k}}
\langle\Psi^{{\bf k},\sigma}_n|
P^t_{m,m'}|\Psi^{{\bf k},\sigma}_n\rangle
\end{equation}
from the minimal condition for $F$, which in this case are independent
of the weights $w_s$.

In the PAW method we obtain the orbital occupations directly from the
PS wave functions $|\tilde\Psi^{{\bf k},\sigma}_n\rangle$ as
\begin{equation}
n^{t,\sigma}_{m,m'}=\frac{1}{I}\sum_{{\bf k},n} \langle\tilde\Psi^{{\bf
k},\sigma}_n|\tilde{P}^t_{m,m'} |\tilde\Psi^{{\bf k},\sigma}_n\rangle
\, ,
\end{equation}
using the pseudo-version
\begin{equation}
\tilde{P}^t_{m,m'}=
\sum_{\Lambda,\Lambda'}
|\tilde{p}_\Lambda\rangle\langle\phi_\Lambda|P^t_{m,m'}
|\phi_{\Lambda'}\rangle\langle\tilde{p}_{\Lambda'}|
+\Delta\tilde{P}^t_{m,m'}
\end{equation}
of the projection operator $P^t_{m,m'}$.  Consistent with the PAW
formalism, the small correction
\begin{equation}
\Delta\tilde{P}^t_{m,m'}=
P^t_{m,m'}
-\sum_{\Lambda,\Lambda'}
|\tilde{p}_\Lambda\rangle\langle\tilde\phi_\Lambda|P^t_{m,m'}
|\tilde\phi_{\Lambda'}\rangle\langle\tilde{p}_{\Lambda'}|
\end{equation}
is ignored in the present calculations because it can be considered the
pseudo-version of the zero operator.

Note that the present choice for the projection operator may not be the
most elegant form. Here we suggest another version using projector
functions $\langle q^t_i|$. The construction is analogous that of the
projector functions in the PAW method, but the purpose is different:\
namely, the goal is to decompose a wave function into local orbitals
such
that
\begin{equation}
\sum_{i,t}|\chi^t_i\rangle\langle q^t_i|\Psi\rangle =|\Psi\rangle
\end{equation}
holds for any wave function $|\Psi\rangle$ that can be represented
exactly as a superposition of the local orbitals $|\chi^t_i\rangle$.
This requirement is fulfilled whenever $\langle
q^t_i|\chi^{t'}_j\rangle=\delta_{i,j}\delta_{t,t'}$.  The projector
functions can be obtained from a set of functions $|f^t_i\rangle$,
which are linearly independent and localized within an augmentation
region as
\begin{equation}
\sum_{j,t'} \langle f^t_i|\chi^{t'}_j\rangle \langle q^{t'}_j|=\langle f^t_i|
\end{equation}
after multiplication with the inverse of $\langle f^t_i|\chi^{t'}_j\rangle$.
If we now use
\begin{equation}
\bar{P}^{t}_{i,j}=|q^t_i\rangle\langle q^{t}_j| \, ,
\end{equation}
we obtain the matrix elements in the representation of localized
orbitals directly as
\begin{equation}
n^{t}_{i,j}=\sum_{k,l,t'',t'''}n^{t'',t'''}_{k,l}\langle\chi^{t''}_k|q^{t}_i\rangle
\langle q^{t}_j|\chi^{t'''}_l\rangle =\sum_{{\bf k},n}\langle\Psi^{{\bf
k}}_n|\bar{P}^{t}_{i,j}|\Psi^{{\bf k}}_n\rangle \, .
\end{equation}
It is now a simple matter to obtain the corresponding pseudo-projection
operator
\begin{equation}
\tilde{\bar{P}}^{t}_{i,j}=\sum_{\Lambda,\Lambda'}
|\tilde{p}_\Lambda\rangle\langle\phi_\Lambda|q^t_i\rangle \langle
q^{t}_j|\phi_{\Lambda'}\rangle\langle\tilde{p}_{\Lambda'}| \, .
\end{equation}

This expression has a number of advantages:\ (i) It is the most general
expression because any operator can be expressed in a fully separable
expression,\cite{seppot} and, unlike Eq.\ (\ref{eq:projector}), it is
not limited to a semi-local form. (ii) The step function in Eq.\
(\ref{eq:projector}) introduces a potential that is discontinuous,
whereas the operator suggested here can be constructed such that the
resulting potential is continuous and differentiable, while still being
fully localized in a well-defined region. (iii) The separable form of
the operator renders the evaluation numerically convenient. In
particular, it is conceivable to evaluate $\Delta\tilde{P}$ explicitly
if necessary, whereas the corresponding operation using the
semi-local projection operator would be computationally prohibitive.
(iv) We can systematically construct projection operators that
specifically act on a well-defined portion of the Hilbert space: By
enlarging the set of functions $|\chi_i\rangle$ to include also
orbitals that should {\em not} be represented in the restricted set,
they are excluded by simply setting the corresponding matrix elements
of the density matrix $N_{i,j}$ to zero.

\subsection{Coulomb and exchange parameters}

The four-center integrals used in the expression of the LDA+U total
energy are defined as
\begin{equation}
\langle\chi^t_i;\chi^t_j|V_{ee}|\chi^t_k;\chi^t_l\rangle =\int {d{\bf
r}_1}\int {d{\bf r}_2} \chi^{t*}_{i}({{\bf r}_1}) \chi^{t*}_{j}({{\bf
r}_2}) v_{ee}({\bf r_1},{\bf r_2}) \chi^t_{k}({{\bf
r}_1})\chi^t_{l}({{\bf r}_2}) \, ,
 \label{eq_u_ijkl_def}
\end{equation}
where $v_{ee}({\bf r},{\bf r'})$ is the screened Coulomb interaction
between two electrons.

If we choose localized $d$-orbitals that are described by an atomic
$d$-wave function $\chi^t_m({\bf r}) =\chi_d(|{\bf r}-{\bf
R}_t|)Y_{d,m}({\bf
  r}-{\bf R}_t)$ with magnetic quantum number $m$ (where $ Y_{d,m}$
are the spherical harmonics for $\ell=2$) and furthermore assume that
the static dielectric function $\epsilon$ is constant in space, we can
exploit the multipole expansion of ${1}/{|{{\bf r}_1}-{{\bf r}_2}|}$:
\begin{equation}
v_{ee}({\bf r}_1,{\bf r}_2)=
\frac{1}{\epsilon \left|{{\bf r}_1}-{{\bf r}_2} \right|}
 = \frac{1}{\epsilon} \sum_{\ell=0}^{\infty}  \frac{4\pi}{2\ell+1}
\frac{r^{\ell}_<}{r^{\ell+1}_>}
 \sum_{m=-\ell}^{+\ell}
  Y_{\ell m}({\bf r_1}) \! Y_{\ell m}^*({\bf r_2}) \, .
\label{eq_multipole_def}
\end{equation}
Here $r_<$ and $r_>$ denote the smallest and largest values of $r_1$ and $r_2$,
respectively.  Under these assumptions we can transform Eq.\
(\ref{eq_u_ijkl_def}) into
\begin{equation}
\langle\chi^t_1;\chi^t_3|V_{ee}|\chi^t_2;\chi^t_4\rangle
 =\sum_{\ell=0}^{\infty}
\frac{4\pi}{2\ell+1} \sum_{m=-\ell}^{+\ell}
\langle\ell_1,m_1|Y_{\ell,m}|\ell_2,m_2\rangle
\langle\ell_3,m_3|Y_{\ell,m}^*|\ell_4,m_4\rangle F^{\ell} \, ,
\label{eq_u_j_clebsch}
\end{equation}
where $(\ell_i,m_i)$ are the angular momenta quantum numbers of
$|\chi^t_i\rangle$, $\langle\ell,m|Y_{\ell'',m''}|\ell,m'\rangle$ the
Gaunt coefficients, and  $F^{\ell}$  the so-called screened Slater's
integrals. Because of the special properties of the Gaunt coefficients,
only $F^0,F^2,$ and $F^4$ contribute to the Coulomb integrals:
\begin{eqnarray}
 F^{\ell}=\frac{1}{\epsilon}
\int_{0}^{\infty}dr_1 \int_{0}^{\infty}
dr_2\ r_1^2 r_2^2
     \chi_d^2(r_1) \chi_d^2(r_2) \frac{r_<^{\ell}}{r_>^{\ell+1}} \, .
\label{slater_integ}
\end{eqnarray}

The parameters $U$ and $J$ are identified with averages of the Coulomb
and exchange integrals, which are related to the Slater integrals $F^0,
F^2,$ and $F^4$ by the properties of Clebsch-Gordan coefficients, Eq.
(\ref{eq_u_j_clebsch}):
\begin{eqnarray}
U&=&\frac{1}{(2\ell+1)^2}\sum_{m,m^\prime}
\langle\chi^t_m;\chi^t_{m'}|V_{ee}|\chi^t_m;\chi^t_{m'}\rangle =F^0 \,
,
\\
J&=&\frac{1}{(2\ell)(2\ell+1)}\sum_{m\ne m^\prime,m^\prime}
\langle\chi^t_m;\chi^t_{m'}|V_{ee}|\chi^t_{m'};\chi^t_m\rangle
=(F^2+F^4)/14 \, ,
\end{eqnarray}
The dielectric constant and therefore the Coulomb and exchange
parameters $U$ and $J$ are not known a priori. Usually, they are
obtained from a constrained DFT calculation.\cite{anisi3} Here, we are
interested in how sensitively the results depend on the choice of
Coulomb parameters, and which Coulomb parameters will provide the best
agreement with reality as probed by optical absorption. Therefore, we
adopt the general form for the four-center integrals as function of the
$U$ and $J$ suggested by the arguments provided above, and perform
calculations for different $U$ values, namely $U =5$ eV$\times I^2$ and
$U = 8$ eV$\times I^2$. In the following we suppose that $I^2 = 1$.
Because the results are fairly insensitive to the exchange parameter
$J$, we have adopted $J=0.95$ eV$\times I^2$ from previous constrained
LDA calculations.\cite{anisi3} The third mandatory relation is obtained
from the work of DeGroot et al.,\cite{groot} who determined the ratio
$F^4/F^2$ for transition-metal oxides to be between 0.62 and 0.63. We
therefore adopt a ratio $F^4/F^2=0.625$.

\subsection{Hamiltonian}

The pseudo-Hamiltonian operator
\begin{equation}
\tilde{H}_\sigma=\tilde{H}_\sigma^{LDA}+\tilde{H}^U_\sigma
 \label{eq_h_ldapu} ,
\end{equation}
which acts on the PS wave functions, is obtained as the derivative of
the total energy functional with respect to the two-center
pseudo-density matrix operator $\tilde\rho=\sum_{n,{\bf
k}}|\tilde\Psi^{{\bf
    k},\sigma}_{n}\rangle f^{{\bf k},\sigma}_n\langle\tilde\Psi^{{\bf
    k},\sigma}_{n}|$.  The non-LDA contribution of the
LDA+U Hamiltonian is then obtained as the product of the derivative
$V^{t,\sigma}_{m,m'}$ of the non-LDA contribution to the total energy and
the projection operator $\tilde{P}$, which is the derivative of the
occupation with respect to the two-center pseudo-density matrix
operator. Thus we obtain
\begin{equation}
\tilde{H}^U_\sigma=\sum_{t,m,m'}\frac{1}{I}\tilde{P}^t_{m,m'}V^{t,\sigma}_{m',m}
\, ,
\label{eq_Hub_def}
\end{equation}
where
\begin{eqnarray}
V^{t,\sigma}_{m_1,m_2}&=&
\sum_{m_3,m_4}
\langle\chi^t_{m_1};\chi^t_{m_3}|V_{ee}|\chi^t_{m_2};\chi^t_{m_4}\rangle
n^{t,-\sigma}_{m_3,m_4}
\nonumber\\
&+&\sum_{m_3,m_4} \bigg[
 \langle\chi^t_{m_1};\chi^t_{m_3}|V_{ee}|\chi^t_{m_2};\chi^t_{m_4}\rangle
-\langle\chi^t_{m_1};\chi^t_{m_3}|V_{ee}|\chi^t_{m_4};\chi^t_{m_2}\rangle
 \bigg] n^{t,\sigma}_{m_3,m_4}
\nonumber\\ &-& \sum_{\sigma'} \left[ U (N^{t,\sigma'}-\frac{1}{2}
\delta_{\sigma,\sigma'}) - \delta_{\sigma,\sigma'} J
(N^{t,\sigma'}-\frac{1}{2}) \right] \delta_{m, m'} \, .
\label{eq_pot_u}
\end{eqnarray}

The LDA contribution of the pseudo-Hamiltonian  has the usual
form:\cite{paw}
\begin{eqnarray}
   {\tilde{H}}^{LDA}_{\sigma}
      = -\frac{\nabla ^{2}}{2} + \widetilde{v} +
    \sum_{\Lambda_1,\Lambda_2}  \!
   \left| \widetilde{p}_{\Lambda_1} \right \rangle
       \left[ \left \langle \phi_{\Lambda_1} \right|
      -\frac{\nabla ^{2}}{2} + v^1  \left|
      \phi_{\Lambda_2}  \right \rangle
      -   \left \langle
       \widetilde{\phi}_{\Lambda_1} \right|
      -\frac{\nabla ^{2}}{2} + \widetilde{v^1} \left|
     \widetilde{\phi}_{\Lambda_2}  \right \rangle \right]
     \left \langle
     \widetilde{p}_{\Lambda_2} \right| \, .
   \label{eq_h_lda}
\end{eqnarray}

For the LDA+U calculation, we first performed a self-consistent LSDA
calculation using the all-electron PAW method, and then used the
self-consistent potential to construct the LSDA Hamiltonian for a large
number of $\bf k$-points in the Brillouin zone. Next, the Hubbard
correction is added to the LSDA Hamiltonian as given by Eq.\
(\ref{eq_h_ldapu}), and the new Hamiltonian iterated until the
occupation numbers $n^{t,\sigma}_{m, m^\prime}$ have converged.  We did
not use the so-called second-variation procedure for self-consistent
LDA+U,\cite{shick} in which the LSDA potential is updated, because it
failed to yield any improvement on calculations similar to
ours.\cite{anisi1}

\section{Ground state of N\lowercase{i}O}
\label{sec:three}
\subsection{LSDA ground state}
\begin{figure}
\begin{center}
\leavevmode
\psfig{figure=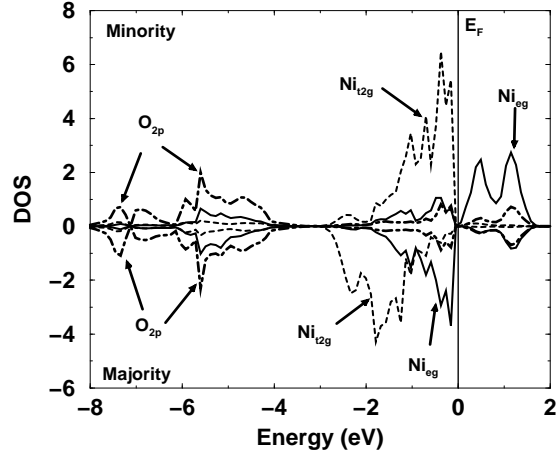,width=85mm}
\end{center}
\caption{
   Atom-resolved antiferromagnetic density of states (DOS, 
  in states per unit cell per eV) of NiO calculated with LSDA.
  The band gap is, with about 0.1 eV, significantly underestimated.
  The spin magnetic moment is 0.95 $\mu_B$.}
\label{lsda_dos}
\end{figure}

The ground state of NiO has been calculated using the PAW method, and
the density of states (DOS) is calculated from the self-consistent PAW
potential using the tetrahedron method for the Brillouin-zone
integration.\cite{ja72}  Figure \ref{lsda_dos} presents the atom-resolved
DOS in the augmentation region.  As can be seen, LSDA produced
an antiferromagnetic insulating ground state with a small band gap. The
LSDA magnetic moment is 0.95 $\mu_B$, and mainly due to the Ni$_{e_g}$
band splitting. This value is much smaller than the experimental value
(1.64--1.90 $\mu_B$, Refs.\ \onlinecite{alperin,cheetham}).  The
calculated band gap of about 0.1 eV is also much smaller than the
experimental one (3.0--4.0 eV). The most interesting feature of our
LSDA DOS is that the $d$-states of Ni dominate the region in the
vicinity of the band gap, and that the top of the valence state is of
Ni$_{e_g}$ type for the first and Ni$_{t_2g}$ for the second spin. This
electronic structure suggests that the band gap is of Mott-Hubbard
type. Hence, this LSDA picture of NiO disagrees completely with
experiment.  It is surprising that the quasi-particle calculation
within the so-called GW approximation, performed by Aryasetiawan and
Gunnarsson,\cite{arya1} produced results qualitatively similar to LSDA
except for an increased band gap of 6 eV and an increased magnetic
moment of 1.6 $\mu_B$. The quantitative change is the reduction of the
O $2p$ bandwidth by almost 1 eV. However, a recent self-consistent
model GW calculation by Massida {\it
  et al.},\cite{massida} in which the dynamic effects were neglected,
produced other results than those of Aryasetiawan and
Gunnarsson.\cite{arya1} It was argued by Massida {\it al.\/} that the
results of the former GW calculation are not quite self-consistent,
presumably because of the additional non-local ad-hoc potential that is
adjusted to the GW calculation in each self-consistent step. The main
difference between the two reported GW model calculations is that the
latter calculation\cite{massida} produced (i) a spreading of the Ni
$d$-states over the entire valence bandwidth, (ii) a vanishing gap
between the O $2p$ and Ni $3d$ and, most importantly, (iii) an
enhancement of O $2p$ states at the valence band maximum. The latter
effect attributes the origin of the band gap mainly to a charge
transfer gap because this gap is now between the O $2p$ and Ni 3$d$
conduction states.  Next, we will show that this latter finding is in
agreement with the results of the LDA+U model.

\subsection{LDA+U ground state}
We have used our implementation of the LDA+U model to determine the
ground-state electronic structure of NiO. Although it is common
practice to use the $U$ extracted from a constrained LDA calculation,
we adopt a different point of view here.  As stated in the introduction
and in agreement with recent results reported in the
literature,\cite{dudarev98} we believe that the value of $U$ extracted
from constrained LDA is not the best possible choice. Therefore we have
determined the electronic structure of NiO for an intermediate $U$ of 5
eV as well as for a larger value of 8 eV.

\begin{figure}
\begin{center}
\leavevmode
\psfig{figure=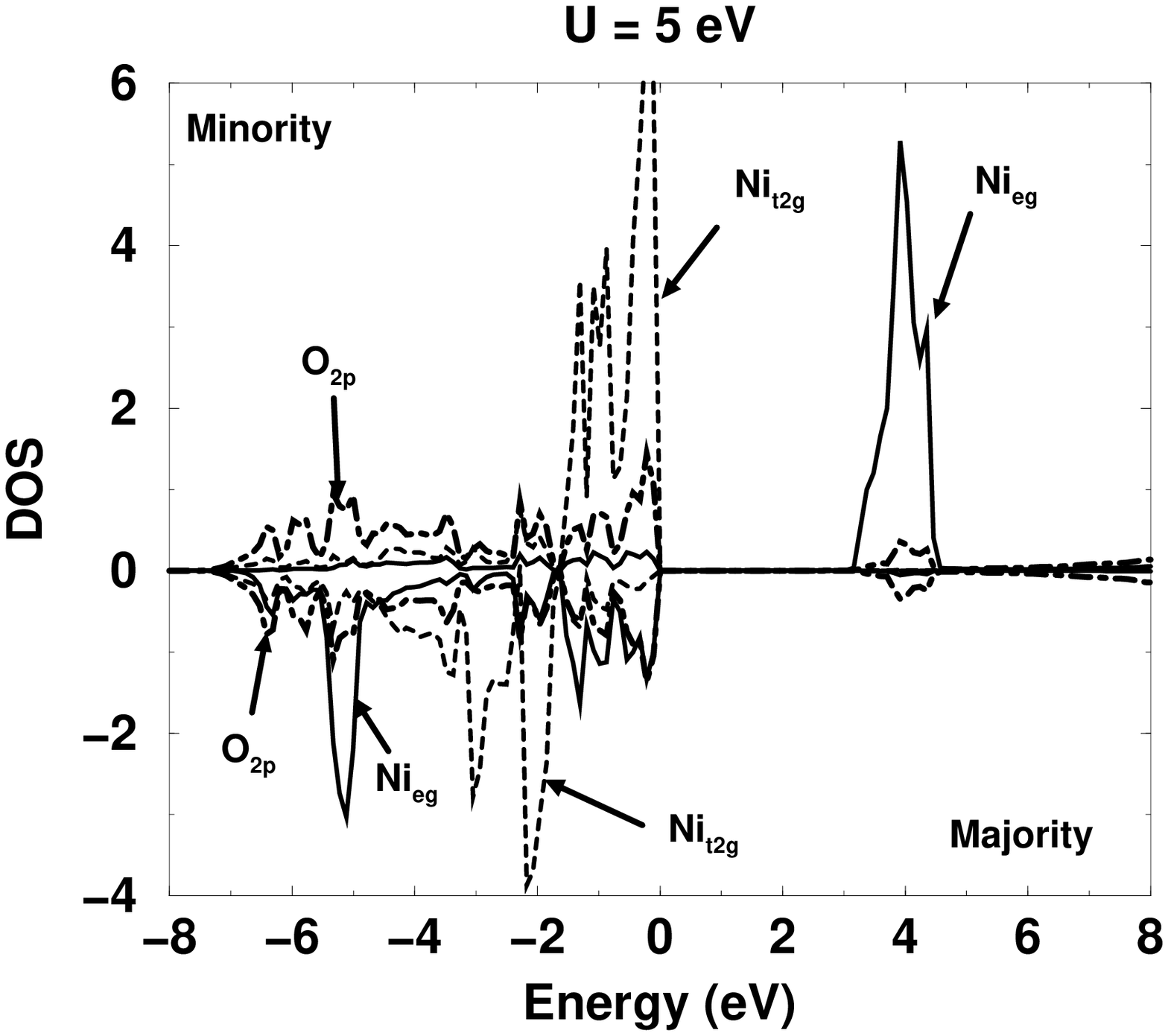,width=85mm}
\psfig{figure=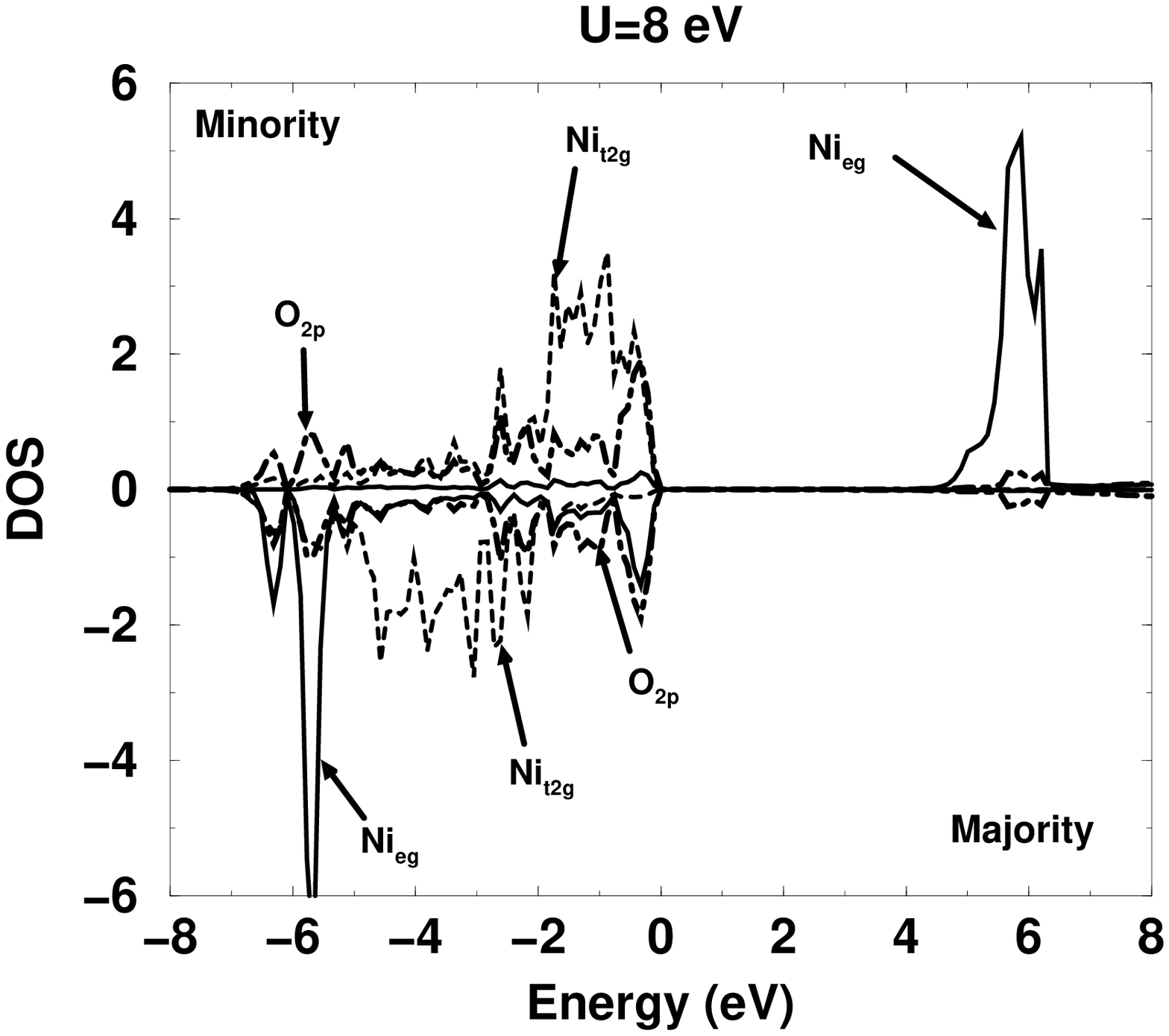,width=85mm}
\end{center}
\caption{
  Atom-resolved density of states (DOS, in states per unit cell per eV)
  of NiO calculated with LDA+U for (a) $U = 5$ eV and (b) $U = 8$ eV . The
  calculated band gap is significantly improved and is $\sim 2.8$ eV ($U =
  5$ eV) and 4.1 eV ($U = 8$ eV).  The antiferromagnetic total magnetic
  moment of 1.73 $\mu_B$ for ($U = 5$ eV) and 1.83 $\mu_B$ for ($U = 8$
  eV) is also in good agreement with experiment. The top of the valence
  state is now O 2$p$-like, thus producing  a mixed
  charge-transfer-type band gap.}
\label{ldapu_dos}
\end{figure}

Figure \ref{ldapu_dos} shows our calculated LDA+U DOS for $U=5$ and 8
eV.  The energy band gap is found to be 2.8 and 4.1 eV, respectively.
The total antiferromagnetic spin moment is 1.73 and 1.83 $\mu_B$,
respectively.  Our DOS obtained for $U=8$ eV agrees well with previous
LDA+U calculations.\cite{anisi1,Surfnio,shick,wei,hugel} However, the
DOS for $U$ = 5 eV is in better agreement with the GW model calculation
of Massida {\it et al}.\cite{massida} The top of valence band is
reinforced by the O $2p$ states, rendering the band gap a mixture of
charge-transfer type and Ni $d \rightarrow d$-like excitations.  In
agreement with previous LDA+U calculations, the spin majority Ni $e_g$
states are pushed towards lower energies, and the energy difference
between $e_g^\uparrow$ and $e_g^\downarrow$ is about 11 eV for $U=8$ eV
and 8.6 eV for $U= 5$ eV. Here again the GW  model yields a value of
about 9 eV for this splitting, in good agreement with our results for
$U$ = 5 eV.

\section{Dielectric function}
\label{sec:four}

For insulators, the imaginary part of the macroscopic dielectric
function is obtained within the random phase approximation (RPA) in the
long wave length limit without local-field effects as\cite{ehrenreich}
\begin{equation}
\epsilon_2(\omega) = \lim_{q \to 0} \sum_{v,c}\sum_{{\bf k}}
   \frac{8 \pi^2}{\Omega q^2} \left| M^{c,{\bf k}}_{v,{\bf k-q}}
   \right|^2  f_{v,{\bf k}} (1-f_{c,{\bf k}})
\delta(E_c^{{\bf k}}-E_v^{{\bf k}} - \hbar\omega) \, .
\label{xxxx}
\end{equation}
Here $M^{c,{\bf k}}_{v,{\bf k-q}}$ are the interband transition matrix
elements,  $f_{v,{\bf k}}$ is the zero-temperature Fermi distribution,
$\Omega$ is the cell volume, $c$ denotes the conduction-band and $v$
the valence-band index. In the case of a local potential, the interband
transition matrix elements are given by
\begin{equation}
\lim_{q \to 0}  M^{c,{\bf k}}_{v,{\bf k-q}} =
\frac{{\bf q}}{\epsilon_{c,{\bf k}}-\epsilon_{v,{\bf k}}}
 \left \langle \Psi_{v}^{\bf k}|{\bf p}|\Psi_{c}^{\bf k}
\right \rangle \, ,
\end{equation}
where the matrix elements $\langle \Psi_v^{\bf k} | {\bf
  p}| \Psi_c^{\bf k}  \rangle $ are calculated using the
PAW crystal wave functions $\Psi^{\bf k}$ described by Eq.\
(\ref{eq_psi_paw}):
\begin{equation}
  \langle\Psi_{v}^{\bf k}|{\bf p}|\Psi_{c}^{\bf k}\rangle  =
  \langle\tilde\Psi_v^{\bf k}|{\bf p}|\tilde\Psi_c^{\bf k}\rangle
  + \sum_{\Lambda,\Lambda^\prime}
  \langle\tilde{\Psi}_{v}^{\bf k}|\tilde{p}_\Lambda\rangle
  \left[
  \langle\phi_\Lambda|{\bf p}|\phi_{\Lambda'}\rangle
  -
  \langle\tilde\phi_{\Lambda}|{\bf p}|\tilde\phi_{\Lambda'}\rangle
  \right]
  \langle\tilde{p}_{\Lambda'}|\tilde\Psi_{c}^{\bf k}\rangle \, ,
  \label{matelements}
\end{equation}
In the most general case, i.e.\ where the potential is nonlocal as in
LDA+U, a nonlocal contribution has to be added to the interband
transition matrix elements.\cite{hybertsen} The full derivation is
given in the Appendix by Eq.\ (\ref{eq_mcvk_def}).  For NiO the
nonlocal contribution to the matrix elements is found to be small,
i.e.\ of a few percent.

\begin{figure}
\begin{center}
\leavevmode\psfig{figure=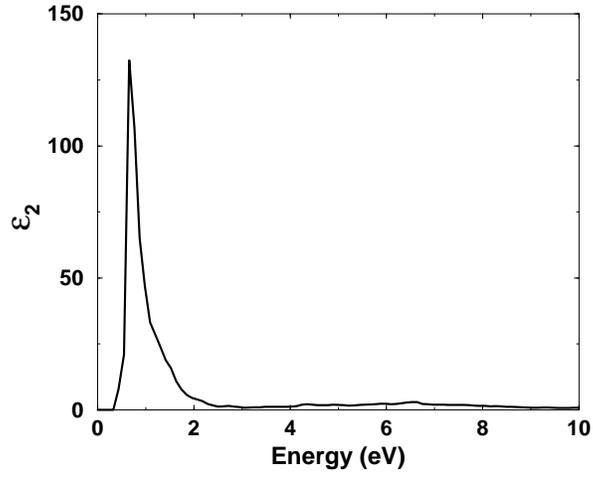,width=85mm}
\end{center}
\caption{
  LSDA calculated imaginary part of the dielectric function
  $\epsilon_2(\omega)$ of NiO.
  Compared to experiment, the optical energy gap is underestimated  and
  the first excitation peak is very intense. This intense peak is due
  to interband transitions from the top of the valence band of O $2p$
  character to the bottom of the conduction band of type Ni $e_g$
  character.  }
\label{lsda_e2}
\end{figure}

\begin{figure}
\begin{center}
\leavevmode
\psfig{figure=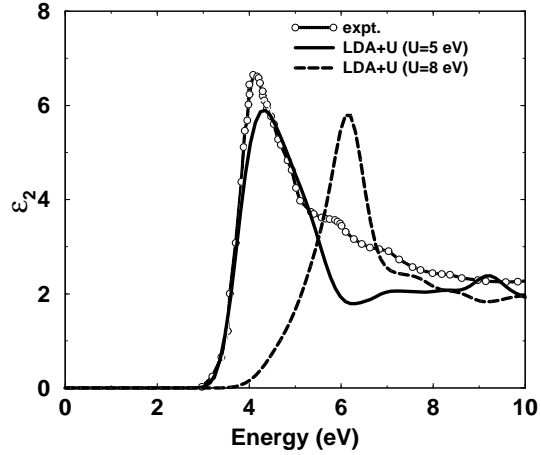,width=85mm}
\end{center}
\caption{
  Calculated LDA+U imaginary part of the dielectric function of NiO
  convoluted with a Gaussian lifetime at half maximum of 0.3 eV for
  $U =5$ eV (solid curve) and for $U = 8$ eV (dashed curve) compared to
  the experimental optical spectrum\protect{\cite{powell}} (dashed-dotted
  curve). The agreement with experiment is much better for $U = 5$ eV.
  Most of the interband transitions giving rise to the first peak are
  from the top of the valence band of O $2$ character to the bottom of
  the conduction band of Ni $e_g$ character.  At higher photon
  energy, the disagreement becomes stronger.}
\label{ldapu_e2}
\end{figure}

Figure \ref{lsda_e2} shows the imaginary part of the dielectric
function calculated within the LSDA. The resulting optical spectrum is
not in agreement with experiment (see Fig.\ \ref{ldapu_e2}). In
particular, the optical gap is considerably underestimated and the
first structure has a much higher intensity compared to experiment.
Conversely,  for $U$ = 5 eV, our calculated imaginary part of the
dielectric function within the LDA+U is in a better agreement with
experiment, as shown in Fig.\ \ref{ldapu_e2}.  The optical band gap and
the oscillator strength of the first excitation peak are in excellent
agreement with experiment. However, at higher photon energies the
agreement with experiment is only qualitative, which is not expected
owing to the mean-field approximation of this simple model. A much
higher value of $U$, i.e.\ 8 eV, produces a much larger optical gap in
contrast to experiment. In agreement with our conclusion that a much
smaller value of $U$ is required to describe NiO, Dudarev and
coworkers\cite{dudarev98} also found that $U=6.2$ eV reproduces the
lattice parameter and the measured  electron-energy loss spectra. It is
too early to draw a definitive conclusion about the excited states of
NiO, as our LDA+U model is a mean-field-like model in which excitonic
effects are not included.  To our knowledge, no calculations of
excitonic effects have so far been attempted. Our results represent the
first investigation of the low-lying excited states of NiO that
considers all the subtleties of chemical bonding {\em and} strong
electron-electron interaction.

\begin{figure}
\begin{center}
\leavevmode\psfig{figure=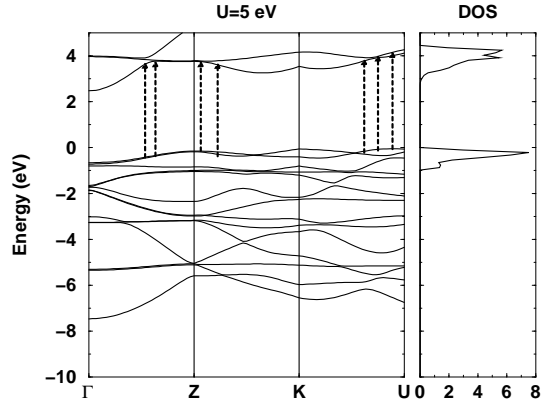,width=85mm}
\end{center}
\caption{
  Left panel:\ band structure of NiO calculated with LDA+U for $U =5$ eV
  along the high-symmetry directions $\Gamma$Z, ZK, and KU.
  Right panel:\ density of states (DOS, in states per unit cell per eV)
  of the initial and final states
  responsible for the first peak of the imaginary part of the
  dielectric function. The interband transitions between parallel
  bands giving rise to the first optical peak are indicated by arrows.  }
\label{bs_dos}
\end{figure}

\begin{figure}
\begin{center}
\leavevmode\psfig{figure=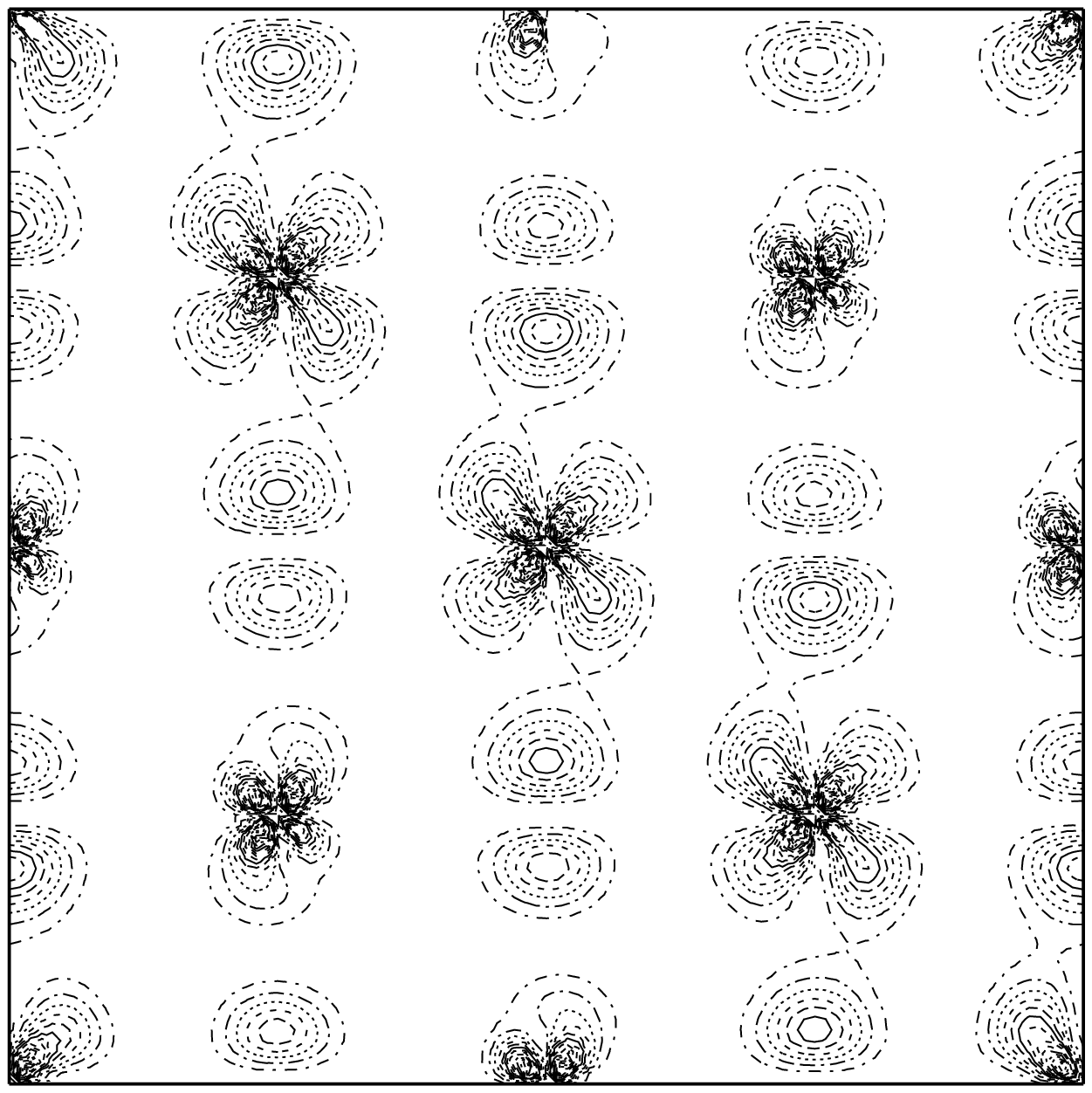,width=85mm}
\leavevmode\psfig{figure=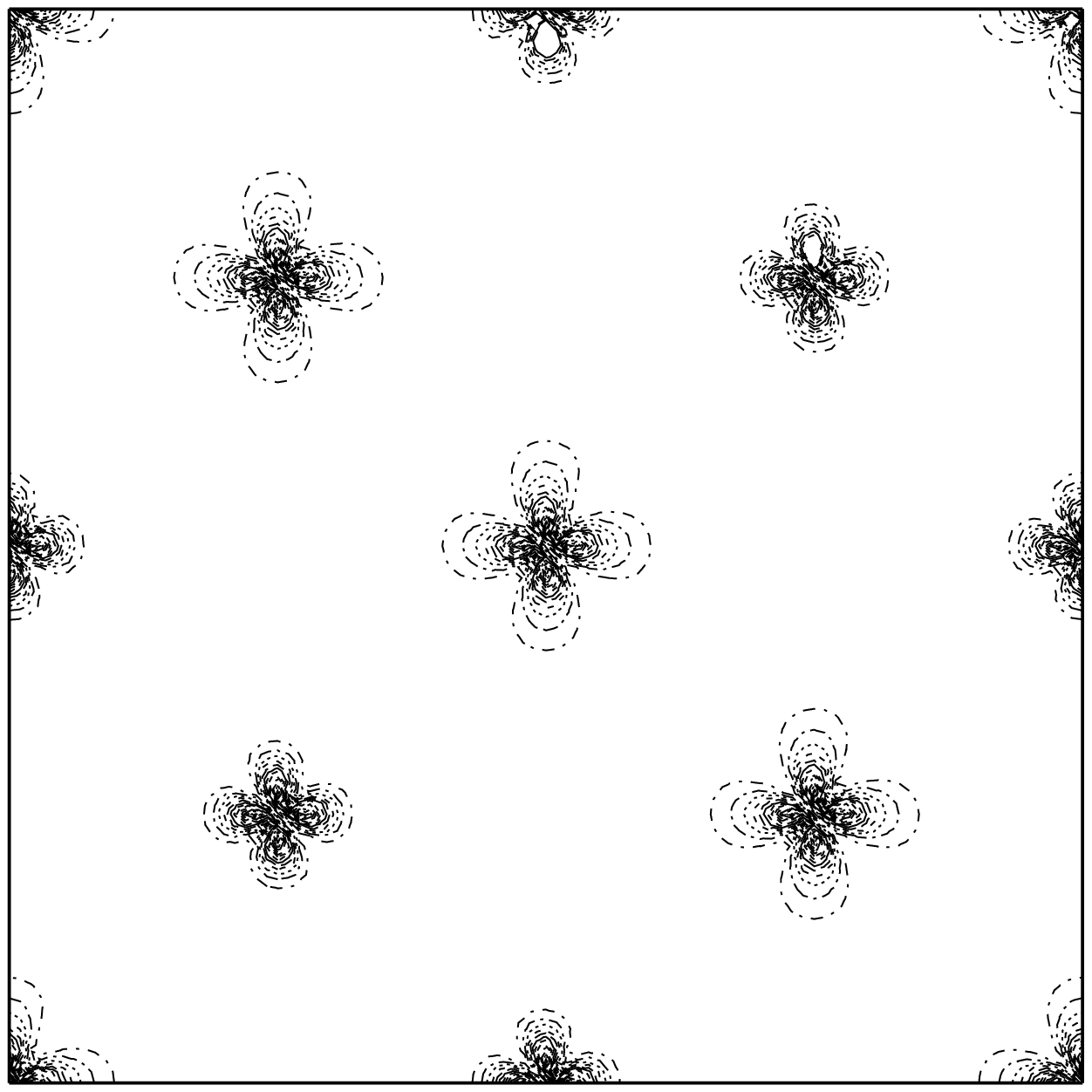,width=85mm}
\end{center}
\caption{
  Initial (top) and final (bottom) charge densities of one spin
  direction at the point ${\bf k}= (\frac{127}{120},
  \frac{127}{360},\frac{\sqrt(3)}{90}) \frac{\pi}{a}$ located between
  the high-symmetry points K and U of the topmost valence band (16)
  and the second lowest conduction band (18), where the optical matrix
  element value is among the largest.  The interband transitions giving
  rise to the first peak of the optical spectrum of NiO are due to
  allowed electronic transition between initial O ${2p}$ and Ni
  ${e_g}$ final states as shown in the plot. }
\label{charge_if}
\end{figure}

The interband transitions are responsible for the first structure in
the optical spectrum of NiO, located between 4.1 and 4.4 eV.  We found
that 40.2\% of the contribution results from the transition from band
15 (second highest occupied band) to band 17 (lowest empty band),
36.2\% from the transition 16$\rightarrow$18, and 15.9\% from the
interband transition 16$\rightarrow$17.  To analyze the character of
the initial and final states of the interband transitions,  the band
structure dispersion along some of the high-symmetry directions is
shown in Fig.\ \ref{bs_dos} together with the DOS of the states that
give rise to the first optical peak.  The arrows between the parallel
bands indicate the interband transitions from the initial to the final
state responsible for the first structure in the optical spectrum.
Figure \ref{charge_if} shows the charge density plot of the initial and
final states $\Psi^{{\bf k},\sigma}_n$ for bands 16 (highest occupied
band) and 18 (second lowest empty band) at point ${\bf k} =
(\frac{127}{120},\frac{127}{360},\frac{\sqrt{3}}{90}) \frac{\pi}{a}$
located between the high-symmetry points K and U, where the optical
matrix element value is among the largest. It also shows that the
initial state is of mixed  O $2p$ and Ni $3d$ character, whereas the
final state  mainly is of Ni $e_g$ character. The optical transitions
then are  between the O $2p$  and the Ni $3d$ states, resulting in
excitations of the charge-transfer type. Our interpretation differs
from that of Fujimori and Minami,\cite{fuji1} who used a configuration
interaction within the metal-ligand cluster and claimed that the
$d\rightarrow d$ charge-transfer transitions are the origin of
fundamental edge.  The drawback of the cluster calculation is that in
reality the oxygen $2p$ orbitals are delocalized and therefore not well
described in a small cluster.  On the other hand, an earlier,
band-structure-based interpretation by Messick and
coworkers\cite{messick} assigned the peak to one electron interband
transitions associated with Ni $3d$ to the Ni $4s$ states. This
interpretation is not correct either because the Ni $4s$ are far above
the top of the valence states (greater than 6 eV), and only quadrupolar
interband transitions are permitted between the $3d$ and $4s$, which
substantially reduces the peak intensity.

\section{Conclusion}
\label{sec:five}

We have presented a new implementation of LDA+U model based on the PAW
method,\cite{paw} which is an all-electron method without any shape
approximation to the potential or the charge density.  We tested the
method on NiO and obtained results that are in good agreement with
previous LDA+U calculations and a recent GW model
calculation.\cite{massida} In particular, we obtained the correct
antiferromagnetic insulating ground state of NiO.

We discussed the results in terms of the strength of the Hubbard
interaction $U$.  The optimum value of $U$ has been determined by
comparison with the experimental dielectric function as well as with
the ground state properties.  We observed a large enhancement of the O
$2p$ character at the top of the valence state, resulting in a
more charge-transfer  than  Ni $d \rightarrow d$ LSDA
-type band gap. The calculated antiferromagnetic moment is in good agreement
with experiment.

Our calculated dielectric function for an intermediate value of $U$,
namely 5 eV, is in good agreement with experiment.  The low-lying,
strong structure in the optical spectrum has been assigned to an
interband transition from O $2p$ states at top of the valence band to
the Ni $e_g$ states at the conduction-band bottom.  Hence the origin of
the first optical peak is due to a charge-transfer excitation.

Our calculation is supported by a recent LDA+U calculation by Dudarev
and coworkers,\cite{dudarev98} who also argue that a much smaller value
of $U$ than the one obtained from constrained LDA calculation is needed
to describe the electron energy loss spectra and the equilibrium
lattice parameter.  It should be interesting to apply this method to
other transition-metal oxides and check the applicability of LSDA+U for
producing excitation energies.

\section{Acknowledgment}
Part of this work was done during our (O.B.\ and M.A.) visit to the
Ohio State University in the summer of 1998. We thank J.\ W.\ Wilkins
and J.\ G.\ LePage for useful discussions. Supercomputer time was
provided by CINES (project gem1101) on the IBM SP2 and by the
Universit\'e Louis Pasteur on the SGI O2000 supercomputer.  \newpage

\appendix
\section{Optical transition matrix elements in LDA+U}
The interband transition matrix elements for a given Hamiltonian are
obtained as follows:
\begin{eqnarray}
M_{v,{\bf k-q}}^{c,{\bf k}}
& = &\langle\Psi_{v,{\bf k-q}} | e^{-i{\bf qr}}|\Psi_{c,{\bf k}}\rangle
 = \frac{\langle\Psi_{v,{\bf k-q}} |(\epsilon_{v,{\bf
      k-q}}-\epsilon_{c,{\bf k}}) e^{-i{\bf qr}}|\Psi_{c,{\bf
      k}}\rangle }{\epsilon_{v,{\bf k-q}}-\epsilon_{c,{\bf k}}}
\nonumber\\
& = &\frac{\langle\Psi_{v,{\bf k-q}} |H e^{-i{\bf qr}}- e^{-i{\bf
      qr}}H|\Psi_{c,{\bf
      k}}\rangle }{\epsilon_{v,{\bf k-q}}-\epsilon_{c,{\bf k}}}
 = \frac{\langle\Psi_{v,{\bf k-q}} |[H, e^{-i{\bf qr}}]_-|\Psi_{c,{\bf
      k}}\rangle }{\epsilon_{v,{\bf k-q}}-\epsilon_{c,{\bf k}}}
      \, .
\end{eqnarray}
The commutator involving of the LDA contribution $H^{\rm LDA}$ to the
Hamiltonian $H=H^{\rm LDA}+H^U$ is
\begin{equation}
[H^{\rm LDA},e^{-i{\bf qr}}]_- =-\frac{1}{2}[\nabla^2,e^{-i{\bf qr}}]_-
=-\frac{1}{2}(-2i{\bf q\nabla}+q^2) =-{\bf qp}+O(q)^2 \, .
\end{equation}
The quadratic and higher-order terms in ${\bf q}$ can be ignored in the
long-wavelength limit appropriate for optical transitions. The
commutator involving the non-LDA part $H^U$ is obtained as
\begin{equation}
[H^{U},e^{-i{\bf qr}}]_-=-i{\bf q}[H^U,{\bf r}]_-+O(q)^2
=-i{\bf q}\sum_{t,m,m'}\frac{V^t_{m,m',\sigma}}{I}[P^t_{m',m},{\bf r}]_-+O(q)^2
\end{equation}
Next we use the relation which holds for the special form of the
projector operator presented in Eq.\ (\ref{eq:projector})
\begin{equation}
P^{t'}_{m'',m'}|\phi_{t,\ell,m,\alpha}\rangle =\theta_{\Omega_t}({\bf
r})|\phi_{t,\ell,m'',\alpha}\rangle
\delta_{t,t'}\delta_{\ell,2}\delta_{m,m'} \, ,
\end{equation}
i.e.\ the projection operator changes the magnetic quantum number of a
specific $d$-like partial wave from $m$ to $m''$ and truncates it
beyond the atomic sphere $\Omega_t$.
Hence we obtain
\begin{equation}
\langle\phi_{\Lambda}|
[ P^t_{m',m},{\bf r}]_-|\phi_{\Lambda'}^t\rangle  =
\langle P^{t\dagger}_{m',m}\phi_{\Lambda}|\theta_{\Omega_t}({\bf r}){\bf r}
|\phi_{\Lambda'}\rangle
-\langle \phi_{\Lambda}|\theta_{\Omega_t}({\bf r}){\bf r}
|P^t_{m,m'}\phi_{\Lambda'}\rangle \, .
\end{equation}

Finally, we obtain the expression of the matrix elements for the
dipole transition, with the PAW LDA+U formalism:
\begin{eqnarray}
\label{eq_mcvk_def}
\lim_{q\to 0}M^{c,{\bf k}}_{v,{\bf k-q}}
& = &
\frac{-{\bf q}}{(\epsilon_{v,{\bf k-q}}-\epsilon_{c,{\bf k}})}
\Bigg\{
\langle\tilde\Psi_{v,{\bf k}}|{\bf p}|\tilde\Psi_{c,{\bf k}}\rangle
+\sum_{\Lambda,\Lambda'}\delta_{t,t'}
\langle\tilde\Psi_{v,{\bf k}}|\tilde{p}_\Lambda\rangle
\bigg[
\langle\phi_{\Lambda}|{\bf p}|\phi_{\Lambda'}\rangle
-
\langle\tilde{\phi}_{\Lambda}|{\bf p}|\tilde{\phi}_{\Lambda'}\rangle
\nonumber\\ &
+ & i\sum_{m,m'}\frac{V_{m,m'}}{I}
(\langle P^{t\dagger}_{m,m'}
\phi_{\Lambda}|\theta_{\Omega_t}({\bf r}){\bf r}|\phi_{\Lambda'}\rangle
-
\langle\phi_{\Lambda}|\theta_{\Omega_t}({\bf r}){\bf r}|P^t_{m,m'}\phi_{\Lambda'}\rangle)
\bigg]
\langle\tilde{p}_{\Lambda'}|\tilde{\Psi}_{c,{\bf k}}\rangle
\Bigg\} \, .
\end{eqnarray}
The difference between wave functions $|\tilde\Psi_{v,{\bf k}}\rangle$
and $|\tilde\Psi_{v,{\bf k-q}}\rangle$ has been ignored because it
only contributes to terms that are proportional to $q^2$, which are ignored
in the long-wavelength limit.


\begin{table}
\caption{
  Magnetic moment and band gap of NiO within LDA, LDA+U ($U=5$ eV)
and LDA+U ($U=8$ eV). The radii of the augmentation regions are 1.7 and
 2.1 a.u.\ for the oxygen and the nickel atoms, respectively.
}
\label{occupation}
\begin{tabular}{rlcccc}
 & & LDA & LDA+U & LDA+U & Expt. \\
 & &     & $U=5$ eV & $U=8$ eV & \\
\hline
m ($\mu_B$) &  & 0.95 & 1.73 & 1.83
& 1.64 (Ref.\ \onlinecite{alperin}) -- 1.9 (Ref.\ \onlinecite{cheetham}) \\
gap (eV) & &  0.1 & 2.8  & 4.1 & 3 -- 4.4 \\
\end{tabular}
\end{table}
\newpage
%
\newpage
%
%
\newpage

%
\newpage

%
\newpage

\end{document}